\documentclass[twocolumn,preprintnumbers,amsmath,amssymb]{revtex4}
\usepackage{graphicx}% Include figure files
\usepackage{dcolumn}% Align table columns on decimal point
\usepackage{bm}% bold math

\usepackage{xcolor}
\usepackage{placeins}
\usepackage{ulem}

\begin{document}

%\preprint{monthly highlight}

\title{High-Resolution Second Harmonic Generation Spectroscopy with Femtosecond Laser Pulses on Excitons in Cu$_2$O}
\author{Johannes Mund$^1$, Dietmar Fr\"ohlich$^1$, Dmitri R. Yakovlev$^{1,2}$ and Manfred Bayer$^{1,2}$}
\affiliation{$^{1}$ Experimentelle Physik 2,
                Technische Universit\"at Dortmund,
                D-44221 Dortmund, Germany}
\affiliation{$^2$ Ioffe Institute,
              Russian Academy of Sciences,
              194021 St. Petersburg, Russia}

\date{\today}

\begin{abstract}
We present a novel spectroscopic technique for second harmonic generation (SHG) using femtosecond laser pulses at 30~kHz repetition rate, which nevertheless provides high spectral resolution limited only by the spectrometer. The potential of this method is demonstrated by applying it to the yellow exciton series of Cu$_2$O. Besides even parity states with $S-$ and $D-$ envelope, we also observe odd parity, $P-$ excitons with linewidths down to 100 $\mu$eV, despite of the broad excitation laser spectrum with a full width at half maximum of 14~meV. The underlying light-matter interaction mechanisms of SHG are elaborated by a group theoretical analysis which allows us to determine the linear and circular polarization dependences, in good agreement with experiment. 
\end{abstract}

%\pacs{78.55.Cr, 78.67.Hc}% , the Physics and Astronomy
%71.70.-d    Level splitting and interactions
%...71.70.Ej    Spin�orbit coupling, Zeeman and Stark splitting, Jahn�Teller effect
%...71.70.Gm    Exchange interactions
%%%%%%%%%%%%%%%%%%%%%%%%%%%%%%%%%%%%%%%%%%%%%%%%%%%%%%%%%%%%%%%%%%%%%%%%%%%%%%%%
%78.20.-e    Optical properties of bulk materials and thin films
%...78.20.Ls    Magnetooptical effects
%%%%%%%%%%%%%%%%%%%%%%%%%%%%%%%%%%%%%%%%%%%%%%%%%%%%%%%%%%%%%%%%%%%%%%%%%%%%%%%%
%78.55.-m    Photoluminescence, properties and materials
%...78.55.Cr    III�V semiconductors
%%%%%%%%%%%%%%%%%%%%%%%%%%%%%%%%%%%%%%%%%%%%%%%%%%%%%%%%%%%%%%%%%%%%%%%%%%%%%%%%
%78.67.-n    Optical properties of low-dimensional, mesoscopic, and nanoscale materials and structures
%...78.67.Hc    Quantum dots
%%%%%%%%%%%%%%%%%%%%%%%%%%%%%%%%%%%%%%%%%%%%%%%%%%%%%%%%%%%%%%%%%%%%%%%%%%%%%%%%


\maketitle

\section{Introduction}

Nonlinear optical techniques, where more than one photon is participating in an elementary excitation process are powerful tools for the investigation of electronic properties of solids \cite{Shen,Boyd,Garmire99,Froehlich94}. Two-photon processes were first treated theoretically almost 90 years ago \cite{Goeppert}, the first experiments on two-photon absorption (TPA) had to wait for the advent of high power lasers more than 50 years ago \cite{Hopfield}. TPA processes are treated in 2nd order perturbation theory which was shown to be equivalent in the tensor formulation to the imaginary part of $\chi^{(3)}$ \cite{Mahr}. Two-photon selection rules were derived from group theory for all crystal symmetries \cite{Inoue}. This approach proved to be for nonlinear spectroscopy of semiconductors advantageous since from the known band structure the symmetry of the exciton eigenfunctions are easily derived which in turn lead to detailed polarization selection rules. The treatment of external peturbations as magnetic and electric field or strain in a first step by group theory gives already helpful insight in the expected mixing of wave functions then to be used in the diagonalization of the appropriate Hamiltonian in order to derive eigenvectors and eigenvalues.

Recently optical second harmonic generation (SHG) has been established as a versatile tool for exciton spectroscopy of semiconductors. At the exciton energies, the SHG intensity is resonantly enhanced, particularly in magnetic and electric fields. Corresponding studies disclosed several field-induced mechanisms of second and third harmonic generation in diamagnetic (GaAs, CdTe, ZnO) and diluted magnetic (Cd, Mn) Te semiconductors \cite{Saenger06a,Saenger06b,Lafrentz13PRL,Lafrentz13,Brunne15}. SHG is a three photon process (two incoming and one outgoing photon). In the tensor formulation it is described as a $\chi^{(2)}$ process. It is a coherent process where two incoming photons are converted to one outgoing photon at twice the energy of the incoming photon \cite{Shen,Boyd}. For use in semiconductor spectroscopy we are interested in resonances for the sum of the two incoming photons.

\begin{figure}[h]
\begin{center}
\includegraphics[width=0.48\textwidth]{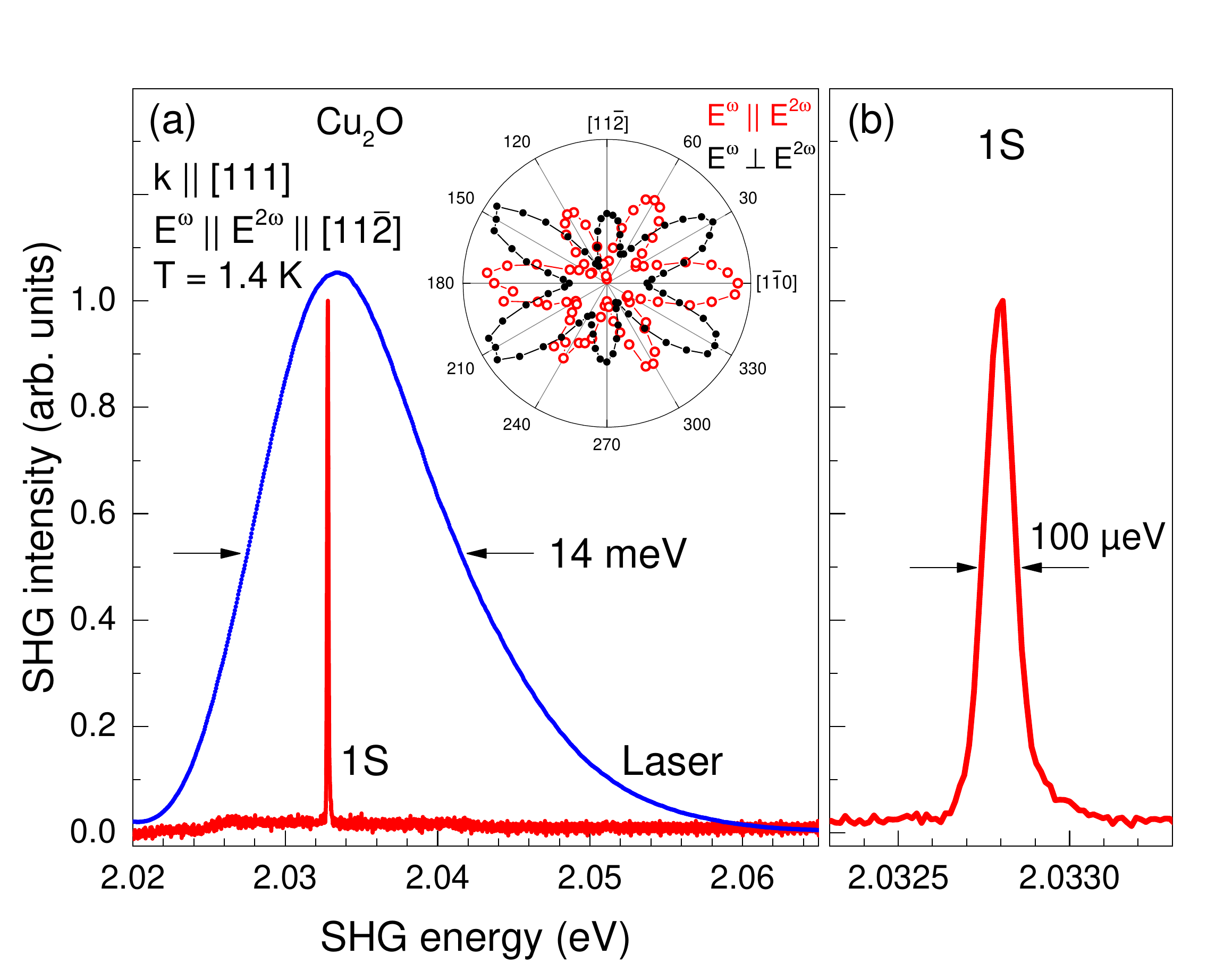}
\caption[Scan]{Exciton SHG spectroscopy in Cu$_2$O  with 200~fs laser pulses. The SHG spectrum of the $1S$ exciton, measured for $\textbf{k}\parallel [111]$, is shown by the red line in panel (a) with its close-up in panel (b). 
The rotational SHG anisotropy at $E_\textrm{1S}=2.0328$~eV is included in panel (a) for two polarization configurations
($\mathbf{E}^\omega \parallel \mathbf{E}^{2\omega}$) and ($\mathbf{E}^\omega \perp \mathbf{E}^{2\omega}$).
The frequency doubled laser spectrum is shown by the blue line.}
\label{pic.1S_SHG+BBO}
\end{center}
\end{figure}

From the point of selection rules for SHG one can consider the mechanism as a twostep process of TPA in absorption and a one photon process in emission. In crystals with inversion symmetry (e.g. $O_h$ crystal symmetry) this process is strictly forbidden for dipole transitions because of parity being a good quantum number. In CuCl, however, $T_d$ symmetry with a p-d mixed valence band, allows TPA and SHG on the same exciton. Indeed in TPA one obtains resonances on the upper polariton and the longitudinal exciton \cite{Froeh_2}, whereas in SHG only resonances on the upper polariton \cite{Hau} are visible.

We show, that in crystals with inversion symmetry (Cu$_2$O, $O_h$ symmetry) one can achieve SHG by replacing one of the odd-parity dipole operators by an even-parity quadrupole operator. We derive polarization selection rules for linearly as well as circularly polarized light. We show, that resonances of even-parity (S/D) as well as odd-parity P excitons can be detected, depending whether the quadrupole operator is applied for the emission of the outgoing photon (leads to even-parity excitons) or for one of the ingoing photons (leads to odd-parity excitons).

In our previous SHG experiments \cite{Saenger06a,Saenger06b,Lafrentz13PRL,Lafrentz13,Brunne15}, we used 7 ns pulses with pulse energy up to 30~mJ, emitted at a low repetition rate of 10~Hz from an optical parametric oscillator pumped by a Nd:YAG laser. The spectral linewidth was limited, however, to about 1~meV. In the present paper on our novel method of SHG specroscopy we show that within the spectral width of the femto-second pulses (about 10~meV) we record spectra on a multi-channel detector behind a monochromator without need for laser tuning with a spectral resolution of 100~$\mu$eV. For larger spectral regions one has to set the energy of the fs laser accordingly. As an example of the potential of our new method of SHG spectroscopy, we show in Fig.~\ref{pic.1S_SHG+BBO} the 100~$\mu$eV wide $1S$ exciton resonance of Cu$_2$O, which was gained by SHG from 200~fs laser pulses. In the inset we show the characteristic six-fold rotational anisotropy, which will be derived in section \ref{sec.theory}. As discussed by Shen \cite{Shen}, efficient SHG with ultrashort pulses requires phase matching. As seen from the $1S$ exciton resonance in CuCl \cite{Froeh_2,Hau} phase matching is always achieved on the upper polariton branch of the resonance. Due to the high repetition (30~kHz) as compared to recording with 10~Hz for each data point in the setup used in our previous experiments, we gain besides the factor of 10 in spectral resolution at least two orders in acquisition time.

The application of this method to Cu$_2$O crystals of different orientation and thickness provides comprehensive insight into the yellow exciton series: In the spectra we see features that can be assigned to $S-$, $P-$, and $D-$excitons with principle quantum numbers up to $n$ = 9, not so far achieved for the even parity states. Moreover, odd-parity exciton SHG had not been detected so far. These results ask for a detailed analysis of two-photon excitation and one-photon emission processes of Cu$_2$O excitons in SHG. The analysis gives also for different crystalline orientations the polarization dependences for linearly and circularly polarized light, in good agreement with the experimental results. The results complement the recent observation of $P$-excitons up to $n=25$ in the Rydberg regime~\cite{Nature}.

The paper is organized as follows: In Sec.~\ref{sec.exp_setup} details of the experimental setup are given. A demonstration of the method with experimental results of clearly resolved resonances of the yellow exciton series up to $n$ = 9 and a resonance of the $1S$ exciton of the green series are given in Sec.~\ref{sec.results}. In Sec.~\ref{sec.theory} we derive in detail the polarization dependences for linearly and circularly polarized light. The resulting rotational anisotropy diagrams are consistent with the experimental results. In the Conclusions we summarize the results and give an outlook for promising applications of the presented new technique.

\section{Experimentals}
\label{sec.exp_setup}

SHG spectroscopy is performed on natural Cu$_2$O semiconductor crystals with high optical quality, as previously demonstrated by high resolution studies of the 1S yellow exciton using a single frequency laser: In absorption, the linewidth of the quadrupole allowed orthoexciton is about a $\mu$eV and that of the paraexciton, mixed with the orthoexciton 
by a magnetic field, is less then 100~neV. \cite{Dasbach_1,Nature}. Crystals of various crystallographic orientations are available: $[001]$, $[1\bar{1}0]$, $[111]$, and $[11\bar{2}]$. The samples are mounted strain-free \cite{Dasbach_1} in suprafluid helium at a temperature of $T=1.4$~K. 

\begin{figure}[h]
\begin{center}
\includegraphics[width=0.48\textwidth]{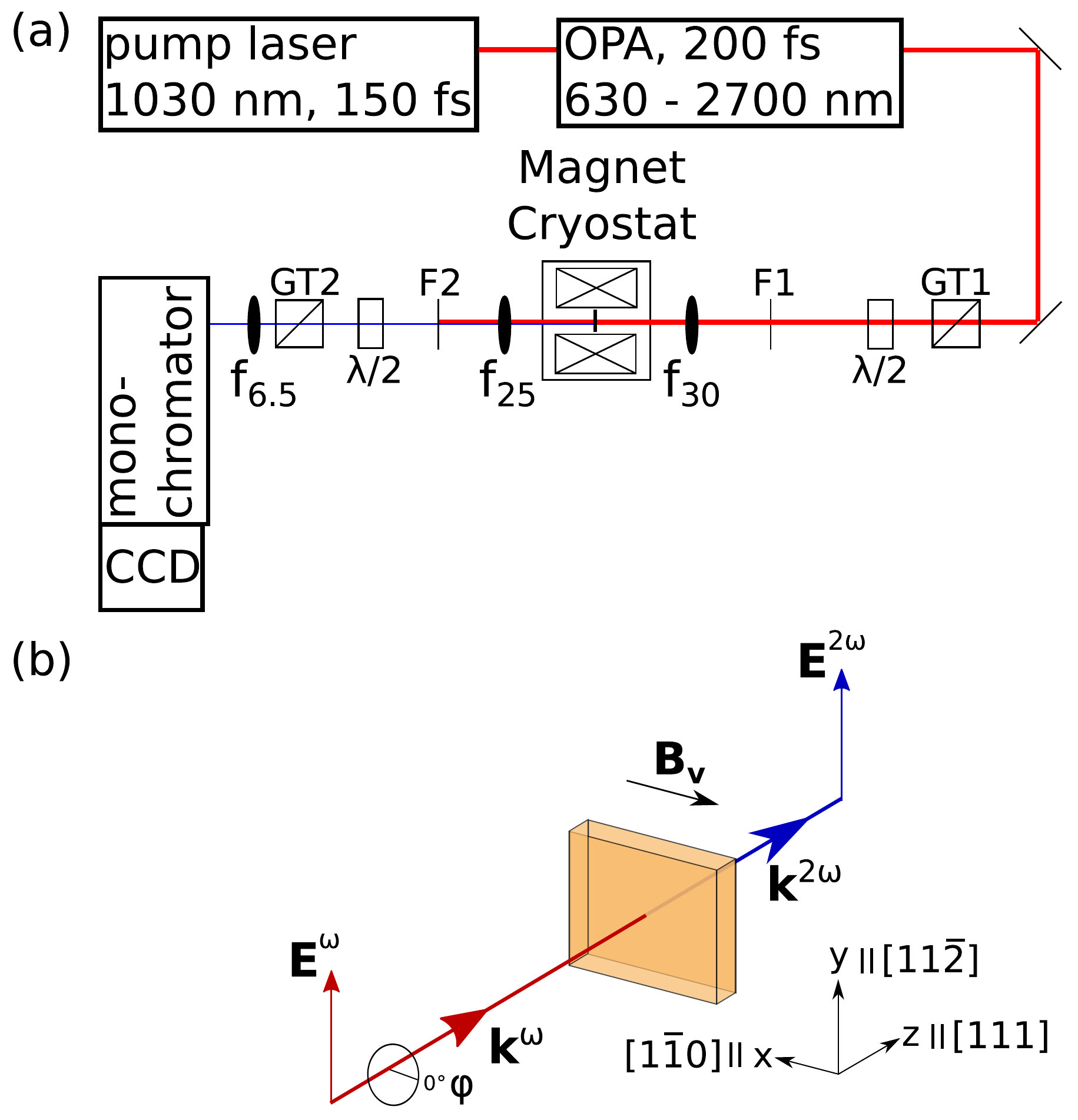}
\caption[Scan]{Upper part: setup for SHG spectroscopy: F - color filter, f$_\textrm{xx}$ - lens with xx-cm focal length, GT - Glan Thompson linear polarizer. Lower part: experimental geometry for Cu$_2$O crystal with $[111]$ orientation, parallel to the optical axis (light vector $\textbf{k}\parallel[111]$). Applied light polarizations $\mathbf{E}^\omega(0^\circ)\parallel[\bar{1}10]$ and $\mathbf{E}^\omega(90^\circ)\parallel[11\bar{2}]$.}
\label{pic.setup}
\end{center}
\end{figure}

The experimental setup is shown in Fig.~\ref{pic.setup}. The exciting laser (Light Conversion) consists of an optical parametric amplifier (OPA, Orpheus) pumped by a pulsed laser (Pharos).  The OPA can be tuned in the spectral range of interest, supplying pulses of about 200~fs duration with a full width at half maximum (FWHM) of $\approx 10$~meV. The energy per pulse can be up to 6~$\mu$J at a repetition rate of 30~kHz. The laser radiation with frequency $\omega$ and wave vector $\mathbf{k}^\omega$ hits the sample normal to its surface and is focused there into a spot of about 100~$\mu$m diameter.

With the Glan Thompson polarizer GT1 and the half-wave plates WP1 and WP2, the linear polarization of the incoming and outgoing light can be varied continuously and independently around the optical axis. One can thus measure for any fixed polarization of $\mathbf{E}^\omega$ or $\mathbf{E}^{2\omega}$ the angular dependence of the other polarization. Of special interest are the rotational anisotropies of the SHG signals for either parallel ($\mathbf{E}^\omega \parallel \mathbf{E}^{2\omega}$) or crossed ($\mathbf{E}^\omega \perp \mathbf{E}^{2\omega}$) polarizations of the laser and the SHG. For experiments with circularly polarized light, the half-wave plates are replaced by quarter-wave plates. GT2 is set to the optimal polarization of the monochromator, providing highest throughput. The long pass filter F1 prevents SHG from optical elements to enter the monochromator and the short pass filter F2 cuts off the infrared pump light. 

The detection of the SHG signal with frequency $2\omega$ and wave vector $\mathbf{k}^{2\omega}=2\mathbf{k}^\omega$ is done by a 0.5-meter monochromator (Acton, Roper Scientific) with a 1800 grooves/mm grating in combination with a Si charge-coupled device (CCD) camera (matrix with $1340 \times 400$ pixels, pixel size 20~$\mu$m) cooled by liquid nitrogen. The overall spectral resolution of the detection system at photon energies of 2~eV is about 100~$\mu$eV. For measurements in a magnetic field, we use a split-coil magnet (Oxford Instruments) allowing field strength up to 10~T.

\section{Experimental Results}
\label{sec.results}

%\subsection{SHG with high spectral resolution}
%\label{ }

As outlined in the introduction and shown in Fig.~\ref{pic.1S_SHG+BBO}, our new method allows us to resolve narrow exciton resonances in SHG spectroscopy, despite the spectrally broad excitation with a femtosecond laser. We excite the $1S$ exciton resonance of Cu$_2$O at $E_\textrm{1S}=2.032775$~eV, determined with a single frequency laser \cite{Froeh_2006}. The center of the laser emission spectrum is accordingly set to 1.017~eV, i.e. roughly at half the $1S$ exciton energy, which is below the sensitivity range of the Si-detector. The laser spectrum after frequency doubling by a BBO ($\beta$-barium borate) crystal, shown by the blue line in Fig.~\ref{pic.1S_SHG+BBO}(a), is broad with a FWHM of 14~meV.

When sending the fundamental light through the Cu$_2$O crystal instead of BBO, SHG appears at the $1S$ energy and this signal is remarkably narrow, see the red line in Fig.~\ref{pic.1S_SHG+BBO}(a) and its close-up in Fig.~\ref{pic.1S_SHG+BBO}(b). The width of the resonance is 100~$\mu$eV, limited by the resolution of the monochromator. The rotational anisotropy, distinguishing coherent SHG processes from photoluminescence after two-photon absorption, is shown in Fig.~\ref{pic.1S_SHG+BBO}(a). Both for parallel ($\mathbf{E}^\omega \parallel \mathbf{E}^{2\omega}$) and crossed ($\mathbf{E}^\omega \perp \mathbf{E}^{2\omega}$) polarizations, the SHG pattern shows a six-fold symmetry pattern.

Due to the observed narrow SHG $1S$ exciton resonance it is appealing to study also the excited states of the yellow exciton series. Due to the exciton Rydberg energy of 86~meV they are located outside of the spectral  range addressed with the fundamental laser energy setting so far. Recent absorption studies with frequency-stabilized lasers had allowed us to follow the series of dipole allowed, odd-parity $P-$excitons up to the principal quantum number $n$ = 25. Furthermore, earlier two-photon absorption studies demonstrated mixed $S-$ and $D-$exciton states of even symmetry up to $n$ = 7 \cite{Uihlein2,Matsumoto, Heck, Schweiner}.  

\begin{figure*}[htpb]
\begin{center}
\includegraphics[height=0.5\textwidth, width=\textwidth]{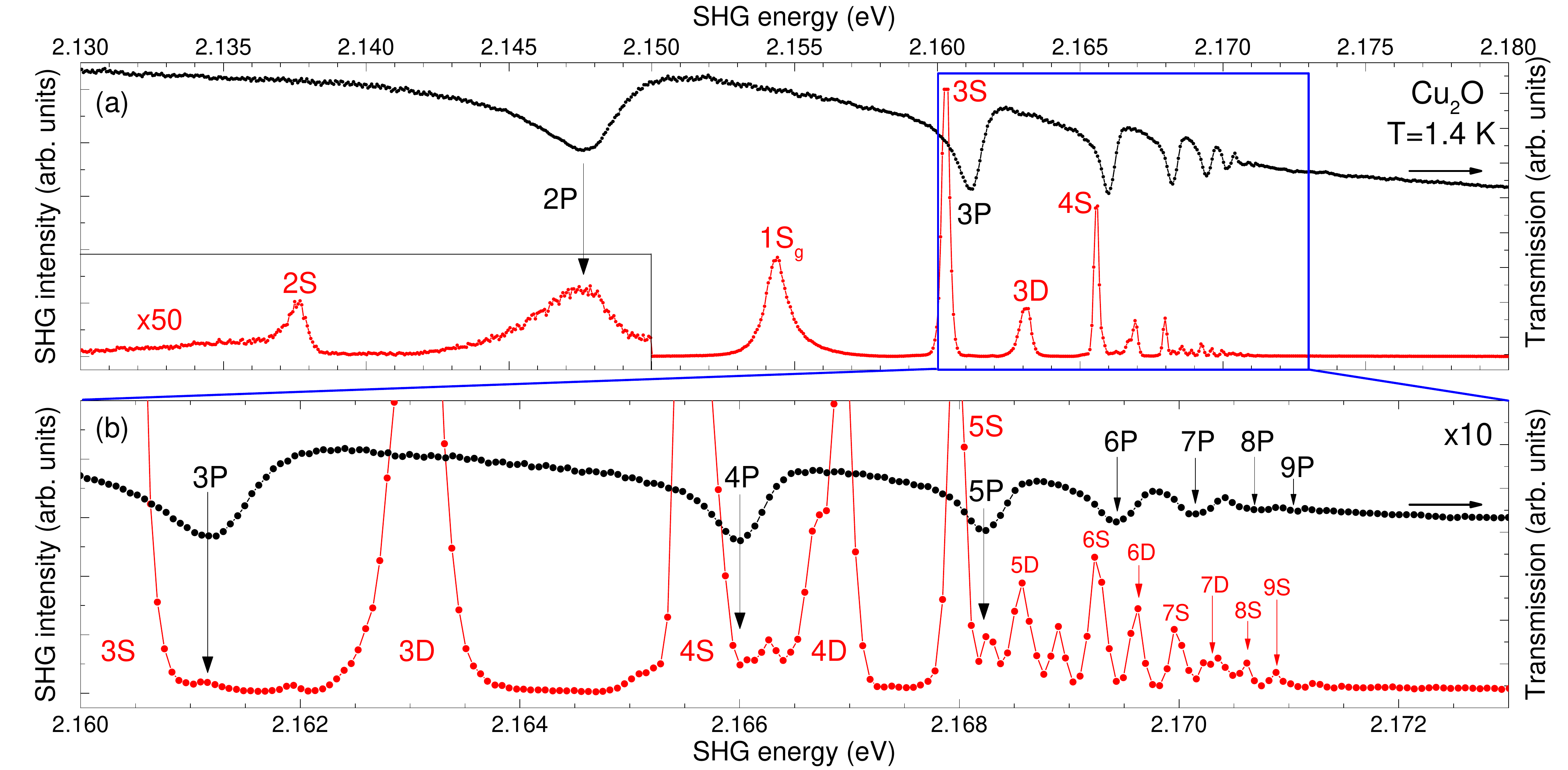}
\caption[Scan]{SHG spectrum (red) of Cu$_2$O  measured in the $\mathbf{E}^\omega \parallel \mathbf{E}^{2\omega} \parallel [11\bar{2}]$ configuration with the photon energy of the fundamental laser set to 1.082~eV and resonant white light transmission spectrum (black), both for $\textbf{k} \parallel [111]$. (a) Energy range from the $n$~=~2 exciton up to more than the band gap of 2.17208~eV \cite{Nature}. Left inset shows the $2S$ and $2P$ resonances, enlarged by a factor of 50. (b) Close-up of the SHG spectrum with an enlargement factor of 10.}
\label{pic.higher_states_[111]_SHG+WL}
\end{center}
\end{figure*}

Corresponding data are shown in Fig.~\ref{pic.higher_states_[111]_SHG+WL}, where we compare a one-photon transmission trace (black) with the SHG spectrum (red) of a thin Cu$_2$O sample (30~$\mu$m thickness), measured in $\textbf{k} \parallel  [111]$ orientation. Here the central photon energy of the fundamental laser was shifted to 1.082~eV. By use of a white light lamp the transmission of the odd-parity $P-$excitons from $n$ = 2 to 9 are observable. In the SHG spectrum we detect $S-$excitons (from $n$ = 2 to 9) and $D-$excitons (from $n$ = 3 to 7, representing in fact mixed $S-D$ states), having even parity. Remarkably, even though we cannot enter the Rydberg regime with $n>9$ exciton states, these data represent an extension beyond previously detected maximal values of principle quantum numbers for the even states.

The SHG spectrum contains even more resonances: Surprisingly, we observe the $2P$ exciton and indications of the $3P$ and $5P$ states. Odd-parity excitons had not been reported in SHG before. This observation asks for a detailed analysis of the underlying light-matter interaction mechanisms that will be given below.

At energy of 2.1544~eV, also the $1S_\textrm{g}$ exciton belonging to the green series is observed in the SHG spectrum. In literature the origin of this state has been debated and assigned to either the $1S_\textrm{g}$ state or to the $2S$ state of the yellow series. We follow here the latest assignment according to  Ref.~\onlinecite{Schweiner}.

\section{Theory}
\label{sec.theory}

SHG is a coherent process composed of two stages: the excitation by two photons involving the same selection rules as in two-photon absorption (TPA) \cite{Inoue} and the one-photon emission.  Both processes need to be allowed to lead to a SHG signal. Here we present model considerations based on a group theory analysis of the symmetries of the exciton states in Cu$_2$O which do not require any assumptions on the optical susceptibility. This straightforward analysis is possible because of the detailed knowledge about the electronic levels involved in the SHG process. Our goal is to understand the mechanisms by which excitons not only of even parity ($S$ and $D$) but also of odd parity ($P$) can be observed in SHG, involving both electric dipole (ED) and electric quadrupole (EQ) transitions. In particular, we extend the analysis to SHG measured not only with linearly, but also with circularly polarized light.

Beforehand, we summarize briefly the basic features of the band structure of Cu$_2$O with point symmetry $O_h$. The highest valence band (VB) and the lowest conduction band (CB) originate from Cu-orbitals, namely from the $3d$- and $4s-$type atomic orbitals, thus these two bands have both even symmetry, the VB of $\Gamma_7^+$-type and the CB of $\Gamma_6^+$-type. Therefore electric dipole transitions between these states are forbidden. Transitions between these two bands lead to the formation of the yellow exciton series. The combination with the exciton envelope wavefunction in addition to the electron and hole Bloch states, allows fulfillment of the dipole transition selection rule for the $P-$excitons. Here we neglect the mixing of different angular momentum states \cite{Uihlein2,Matsumoto, Heck, Schweiner}. The $S-$ and $D-$excitons can be excited in one-photon-transitions only on the quadrupole level with correspondingly small transition matrix elements \cite{Heck}.

The green exciton series, on the other hand, involves besides the lowest CB the first excited VB to which also $3d$ Cu-orbitals contribute. This band is split from the lowest VB by 130~meV due to spin-orbit interaction. The band symmetry is even with $\Gamma_8^+$, thus electron-hole transitions are as well dipole-forbidden.

\subsection{Group theory of SHG process}

\begin{figure}[h]
\begin{center}
\includegraphics[width=0.4\textwidth]{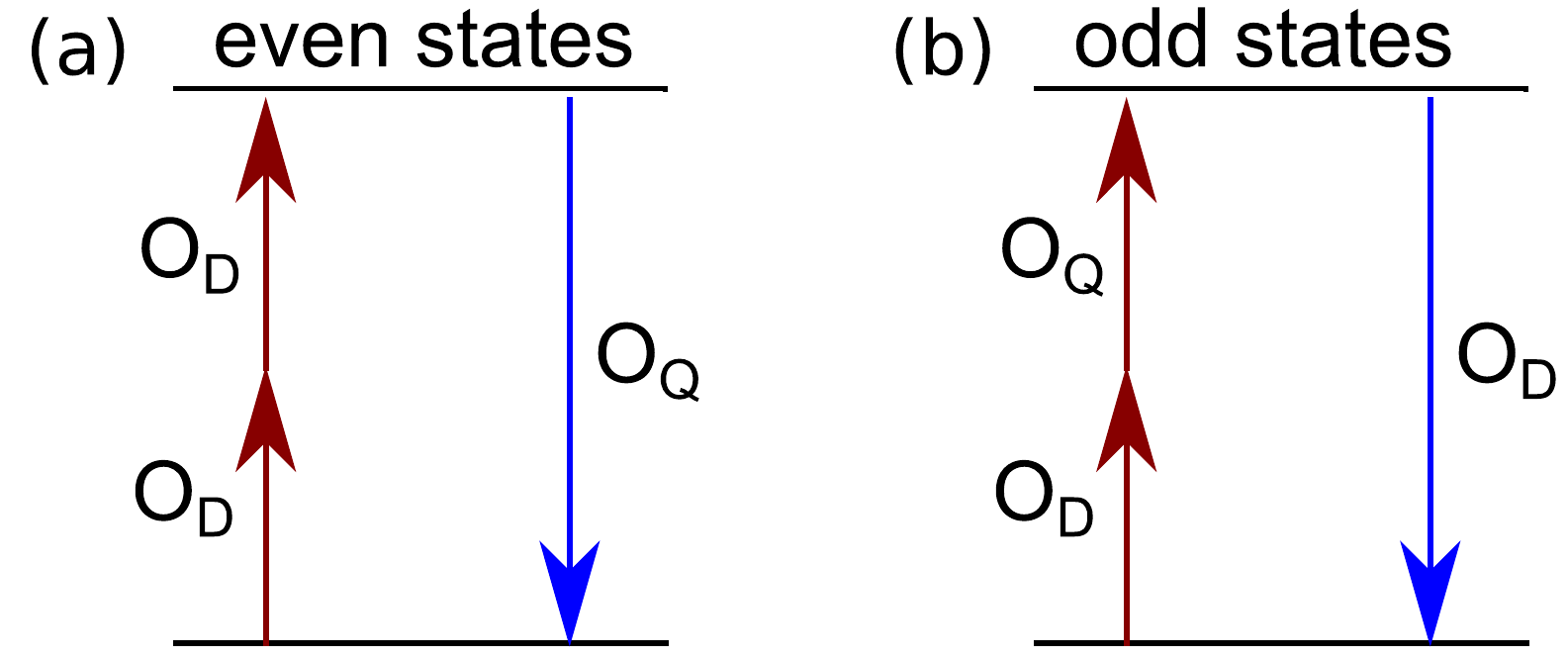}
\caption[Scan]{Sketch of resonant SHG processes for the even and odd-parity exciton states in point group O$_\textrm{h}$. 
}
\label{pic.SHG-process}
\end{center}
\end{figure}

In the point group $O_h$, the dipole operator, $O_\textrm{D}$, transforms as the irreducible representation $\Gamma^-_4$ and the quadrupole operator, $O_\textrm{Q}$,  as $\Gamma^+_5$. With the symmetries of these operators, the required operator combinations for SHG on excitons of different parities in Cu$_2$O can be determined. These combinations are indicated in Fig.~\ref{pic.SHG-process}.

Let us consider first the even exciton SHG (Fig.~\ref{pic.SHG-process}\,a). Applying twice the odd-parity dipole operator $O_\textrm{D}$ ($\Gamma^-_4$ symmetry) for the two incoming photons leads to excitation of even-parity states with possible symmetries $\Gamma^+_1$, $\Gamma^+_3$, $\Gamma^+_4$ and $\Gamma^+_5$, as was first derived for two-photon absorption in Ref. \cite{Inoue}. The subsequent one-photon emission of the even parity exciton states cannot be provided by the dipole operator $O_\textrm{D}$ due to parity reasons. However, this transition is allowed by the quadrupole operator $O_\textrm{Q}$ that can be decomposed in $\Gamma^+_3$ and $\Gamma^+_5$ symmetry. Since the $S$-excitons in Cu$_2$O have $\Gamma^+_5$ symmetry, only the $\Gamma^+_5$ component of the quadrupole operator is relevant for the outgoing photon.

The odd parity excitons in SHG (Fig.~\ref{pic.SHG-process}\,b), on the other hand, can be excited by replacing one of the dipole operators ($\Gamma^-_4$ symmetry) in the two-photon excitation by the quadrupole operator ($\Gamma^+_5$ symmetry) and the quadrupole operator in the emission process by a dipole operator. From the tables of Koster et al. \cite{Koster} the combined two-photon dipole-quadrupole operator can be decomposed into the irreducible representations $\Gamma^-_2$, $\Gamma^-_3$, $\Gamma^-_4$ and $\Gamma^-_5$. Since the decomposition contains the $\Gamma^-_4$ term, dipole emission of a single photon is possible in the SHG process.  

It follows from our symmetry analysis that in Cu$_2$O both even and odd parity exciton states can contribute to SHG. This is in full agreement with our experimental results in Figs.~\ref{pic.1S_SHG+BBO} and \ref{pic.higher_states_[111]_SHG+WL}, where even $S$ and $D$ exciton states and odd $P$ states show up. After having outlined the appearance of the different features, we want to understand the SHG processes in more detail by analyzing their polarization properties.

\subsection{SHG polarization}

\subsubsection{Linearly polarized light}

With use of the Tables of Koster et al. \cite{Koster}, we can derive the polarization dependence of the SHG signals for linearly polarized light as function of the polarization vectors of the incoming pair of photons $\mathbf{E}^\omega(\varphi)$ and the outgoing photon $\mathbf{E}^{2\omega}(\varphi)$ as well as their wavevector $\mathbf{k}$:

\begin{equation}\label{eq.photon_pols}
\mathbf{E}^\omega(\varphi)=\begin{pmatrix}
u(\varphi)\\v(\varphi)\\w(\varphi)
\end{pmatrix},\,
\mathbf{E}^{2\omega}(\varphi)=\begin{pmatrix}
m(\varphi)\\n(\varphi)\\o(\varphi)
\end{pmatrix},\,
\mathbf{k}=\begin{pmatrix}
x\\y\\z
\end{pmatrix} .
\end{equation}
For the excitation of even parity excitons (Fig.~\ref{pic.SHG-process}(a)), the combination of the dipole operators for the two exciting photons is given by:

\begin{equation}\label{eq.O_TPD}
O_\textrm{TPDD}(\mathbf{E}^\omega)=\sqrt{2}\begin{pmatrix}
v(\varphi) w(\varphi)\\u(\varphi) w(\varphi)\\u(\varphi) v(\varphi)
\end{pmatrix}.
\end{equation}
For the outgoing photon, the $\Gamma^+_5$ quadrupole operator is given by:
\begin{equation}\label{eq.O_Q}
O_\textrm{Q}(\mathbf{k}, \mathbf{E}^{2\omega})=\frac{1}{\sqrt{2}}\begin{pmatrix}
y \cdot o(\varphi)+z\cdot n(\varphi)\\z\cdot m(\varphi)+x\cdot o(\varphi)\\x\cdot n(\varphi)+y\cdot m(\varphi)
\end{pmatrix}.
\end{equation}
The intensity of the SHG signal, $I^{2\omega}$, is proportional to the square of the product of the operators for the ingoing and outgoing transitions. The SHG signal of even parity excitons ($S$ and $D$ excitons in Cu$_2$O) as function of the relevant parameters (Eq.~\ref{eq.photon_pols}) is therefore given by:

\begin{equation}\label{eq.SH_Oh_even}
I^{2\omega}_\textrm{even} \propto | O_\textrm{TPDD} \cdot O_\textrm{Q}|^2.
\end{equation}

For the excitation of odd parity states (Fig.~\ref{pic.SHG-process}\,b) we start with the 
$\Gamma^-_4$ term of the combination of the dipole and the quadrupole operator $O_\textrm{TPDQ}$ in the excitation process:
\begin{equation}\label{eq.O_TPDQ}
O_\textrm{TPDQ}(\mathbf{E}^\omega, \mathbf{q}^\omega)=\frac{1}{\sqrt{2}}\begin{pmatrix}
v(\varphi) q_z(\varphi)&+&w(\varphi) q_y(\varphi)\\w(\varphi) q_x(\varphi)&+&u(\varphi) q_z(\varphi)\\u(\varphi) q_y(\varphi)&+&v(\varphi) q_x(\varphi)
\end{pmatrix} .
\end{equation}
The components of the quadrupole polarization vector $\mathbf{q}^\omega$ are calculated from Eq.~\eqref{eq.O_Q} by replacing the components of $\mathbf{E}^{2\omega}$ by the components of $\mathbf{E}^\omega$. For the outgoing photon, we now have to apply the dipole operator:
\begin{equation}\label{eq.O_D}
O_\textrm{D}(\mathbf{E}^{2\omega})=\begin{pmatrix}
m(\varphi)\\n(\varphi)\\o(\varphi)
\end{pmatrix} .
\end{equation}
The SHG signal for odd parity states ($P-$excitons in Cu$_2$O) is thus given by:
\begin{equation}\label{eq.SH_Oh_odd}
I^{2\omega}_\textrm{odd} \propto |O_\textrm{TPDQ} \cdot O_\textrm{D} |^2.
\end{equation}
Note, that we derive here equations for the SHG polarization dependence  only from symmetry considerations of the transition operators. In order to get absolute values of $I^{2\omega}$ or its relative values for various exciton states, additional information on the transition matrix elements would be necessary.

\subsubsection{Circularly polarized light}

Commonly, SHG spectra are excited and analyzed using linear polarization for the exciting laser and the SHG signal. Here we extend SHG to circular polarization. We derive the SHG polarization dependence for circularly polarized light following closely the calculation for two-photon absorption \cite{Froeh_4}. For circularly polarized light only the $k$-direction of the photons is of relevance. It is convenient to introduce for any arbitrary $k$-direction two orthogonal vectors $\mathbf{d}(x,y,z)$ and $\mathbf{e}(x,y,z)$ as basis vectors for the complex polarization vectors $\sigma^+(x,y,z)$ and $\sigma^-(x,y,z)$. 
For a $k$-vector
\begin{equation}
\mathbf{k}=\begin{pmatrix}
x\\y\\z
\end{pmatrix}\\
\end{equation}
we get:
\begin{eqnarray}
\mathbf{d}=\begin{pmatrix}
1-x^2/(1+z)\\\ -x y/(1+z)\\-x
\end{pmatrix}, \,\,\,
\mathbf{e}=\begin{pmatrix}
-x  y/(1+z)\\1-y^2/(1+z)\\-y
\end{pmatrix} .
\end{eqnarray}
From these orthogonal polarization vectors we define the circular polarization vectors:
\begin{eqnarray}
\sigma^+=\frac{\mathbf{d}-i\cdot \mathbf{e}}{\sqrt{2}}, \,\,\,\,
\sigma^-=\frac{\mathbf{d}+i\cdot \mathbf{e}}{\sqrt{2}} .
\label{eq.sigma}
\end{eqnarray}

We consider first even parity excitons shown in Fig.~\ref{pic.SHG-process}(a) and use the operators defined by Eqs.~\eqref{eq.O_TPD} and~\eqref{eq.O_Q}. By inserting the components of right (R) and left (L) circularly polarized photons from Eq.~\eqref{eq.sigma} we get the operators for the even parity states:
\begin{equation}
O_\textrm{TPDD$_\textrm{R}$}(\sigma^+),\,
O_\textrm{TPDD$_\textrm{L}$}(\sigma^-),\,
O_\textrm{Q$_\textrm{R}$}(\sigma^+),\,
O_\textrm{Q$_\textrm{L}$}(\sigma^-) .
\label{  }
\end{equation}
By inserting the circular polarization vectors from Eq.~\eqref{eq.sigma} in Eqs. \eqref{eq.O_TPDQ} and \eqref{eq.O_D} we get the operators for the odd parity states:
\begin{equation}
O_\textrm{TPDQ$_\textrm{R}$}(\mathbf{k}, \sigma^+), \,
O_\textrm{TPDQ$_\textrm{L}$}(\mathbf{k}, \sigma^-), \,
O_\textrm{D$_\textrm{R}$}(\sigma^+), \,
O_\textrm{D$_\textrm{L}$}(\sigma^-) .
\end{equation}

The SHG signals for the different polarization combinations of ingoing and outgoing photons are thus given by:
\begin{eqnarray}
\label{IRR}
I^{2\omega}_\textrm{RR, even} &\propto& | O_\textrm{TPDD$_\textrm{R}$} \cdot O_\textrm{Q$_\textrm{R}$}|^2\\
I^{2\omega}_\textrm{RL, even}&\propto&| O_\textrm{TPDD$_\textrm{R}$} \cdot O_\textrm{Q$_\textrm{L}$}|^2\\
I^{2\omega}_\textrm{RR, odd}&\propto&| O_\textrm{TPDQ$_\textrm{R}$} \cdot O_\textrm{D$_\textrm{R}$}|^2\\
I^{2\omega}_\textrm{RL, odd}&\propto&| O_\textrm{TPDQ$_\textrm{R}$} \cdot O_\textrm{D$_\textrm{L}$}|^2 .
\label{IRL}
\end{eqnarray}
As an example we present in the following details of the calculations for even and odd parity exciton states in the $\textbf{k} \parallel [111]$ orientation. The vectors $\mathbf{k}$, $\sigma^+$ and $\sigma^-$ have the explicit forms:
\begin{eqnarray}
\textbf{k}&=&\frac{1}{\sqrt{3}}\begin{pmatrix}
1\\1\\1
\end{pmatrix}\\
\sigma^+&=&\frac{1}{\sqrt{3}}\begin{pmatrix}
(\sqrt{3}-i)/2\\-(\sqrt{3}+i)/2\\i
\end{pmatrix}\\
\sigma^-&=&\frac{1}{\sqrt{3}}\begin{pmatrix}
(\sqrt{3}+i)/2\\(-\sqrt{3}+i)/2\\-i
\end{pmatrix} .
\end{eqnarray}
The resulting relative intensities of the SHG signals for circularly and linearly polarized photons are given in Table~\ref{Table1}.
\begin{table}[htbp]
\begin{center}
\caption{Relative intensities for linearly and circularly polarized SHG signals in Cu$_2$O for $\mathbf{k} \parallel [111]$.}
\label{tab.osc_strengths}
\label{Table1}
\begin{tabular}{|c|c|c|c|c|c|}
\hline
in/out & lin/lin & $\sigma^+$/$\sigma^+$ & $\sigma^-$/$\sigma^-$ & $\sigma^+$/$\sigma^-$ & $\sigma^-$/$\sigma^+$\\ \hline
O$_\textrm{h}$ even & 1/18 & 0 & 0 & 1/9 & 1/9\\
O$_\textrm{h}$ odd & 2/9 & 0 & 0 & 4/9 & 4/9\\ \hline
\end{tabular}
\end{center}
\end{table}

Four configurations of circularly polarized light, labeled by the polarization of "laser/SHG", can be examined. Obviously, SHG is forbidden for the co-circular polarization configurations $\sigma^+/\sigma^+$ and $\sigma^-/\sigma^-$, but allowed for the crossed-circular polarizations $\sigma^+/\sigma^-$ and $\sigma^-/\sigma^+$. For the allowed configurations the intensity of the circularly polarized SHG is by a factor of 2 larger than for linearly polarized light. This is valid for both, even and odd parity exciton states.

\subsubsection{SHG polarization in experiment}

\begin{figure}[h]
\begin{center}
\includegraphics[width=0.5\textwidth]{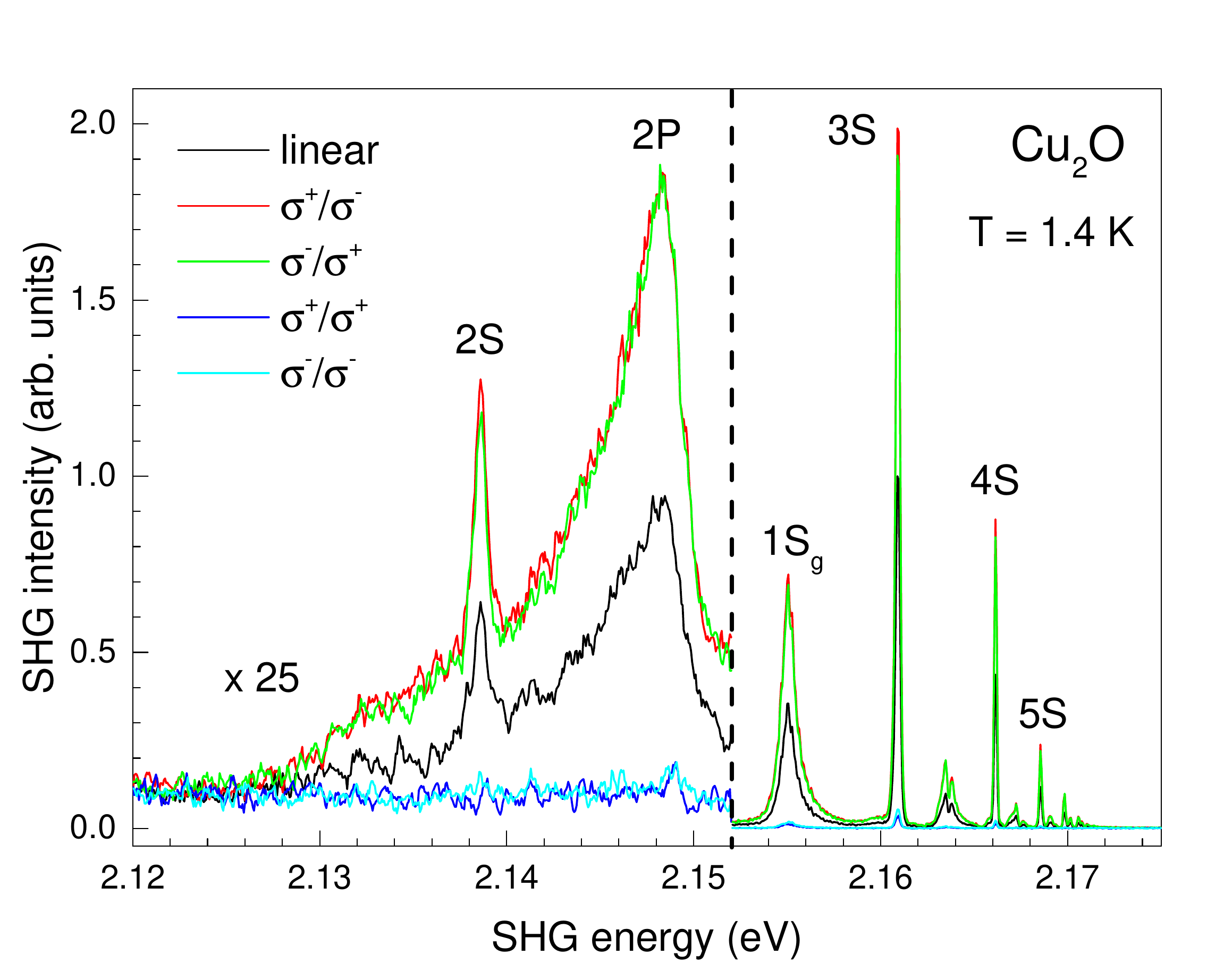}
\caption[Scan]{Polarization dependence of SHG spectra in Cu$_2$O for $\mathbf{k} \parallel [111]$. The linear polarization setting is $\mathbf{E}^\omega \parallel \mathbf{E}^{2\omega} \parallel [11\bar{2}]$. For the low energy side the central laser energy is set to 1.073~eV and for the high energy side to 1.082~eV. The spectra are normalized to the maximum SHG intensity in linear polarization (at $2P$ for left side and at $3S$ for right side). Note the multiplication factor of 25 for the weaker signals on the left side.}
\label{pic.higher_111_direction}
\end{center}
\end{figure}

Here we test the predicted dependences of the SHG intensity on the light polarization. Corresponding SHG spectra for circularly polarized excitation and the comparison to linear polarization are presented in Fig.~\ref{pic.higher_111_direction} for $\textbf{k}\parallel[111]$ orientation. The signals are normalized to the linearly polarized SHG intensity of the $2P$ and $3S$ states for the low- and high-energy parts, respectively. 

SHG signals are detected in cross-circular polarizations ($\sigma^+/\sigma^-$ and $\sigma^-/\sigma^+$) for which they are found to be twice as strong as for linearly polarized light. On the other hand, for the co-circular polarizations ($\sigma^+/\sigma^+$ and $\sigma^-/\sigma^-$) the SHG signals are absent. These findings are in excellent agreement with the above predictions.

\subsection{Rotaional anisotropies}

\subsubsection{Calculated anisotropies}

Next we present details of the SHG rotational anisotropies for the two orientations relevant for the experimental results presented below. The directions for $\mathbf{k}$, $\textbf{E}^\omega(\varphi)$ and $\textbf{E}^{2\omega}(\varphi)$ in Eq.~\eqref{eq.SH_Oh_odd} for the odd exciton state in the $\textbf{k}\parallel [111]$ configuration are:
\begin{eqnarray}
\mathbf{k}&=&\frac{1}{\sqrt{3}}\begin{pmatrix}
1\\1\\1
\end{pmatrix}\\
\textbf{E}^\omega(\varphi)=\textbf{E}^{2\omega}(\varphi)&=&\frac{1}{\sqrt{6}}\begin{pmatrix}
\textrm{cos}(\varphi)-\sqrt{3}\,\textrm{sin}(\varphi)\\\textrm{cos}(\varphi)+\sqrt{3}\,\textrm{sin}(\varphi)\\-2\,\textrm{cos}(\varphi)
\end{pmatrix}
\end{eqnarray}
and for the even exciton states in the $\textbf{k} \parallel [11\bar{2}]$ orientation they are:
\begin{eqnarray}
\mathbf{k}&=&\frac{1}{\sqrt{6}}\begin{pmatrix}
1\\1\\-2
\end{pmatrix}\\
\textbf{E}^\omega(\varphi)=\textbf{E}^{2\omega}(\varphi)&=&\frac{1}{\sqrt{3}}\begin{pmatrix}
\textrm{cos}(\varphi)+\sqrt{\frac{3}{2}}\,\textrm{sin}(\varphi)\\\textrm{cos}(\varphi)-\sqrt{\frac{3}{2}}\,\textrm{sin}(\varphi)\\\textrm{cos}(\varphi)
\end{pmatrix}.
\end{eqnarray}

\begin{figure}[h]
\begin{center}
\includegraphics[width=0.48\textwidth]{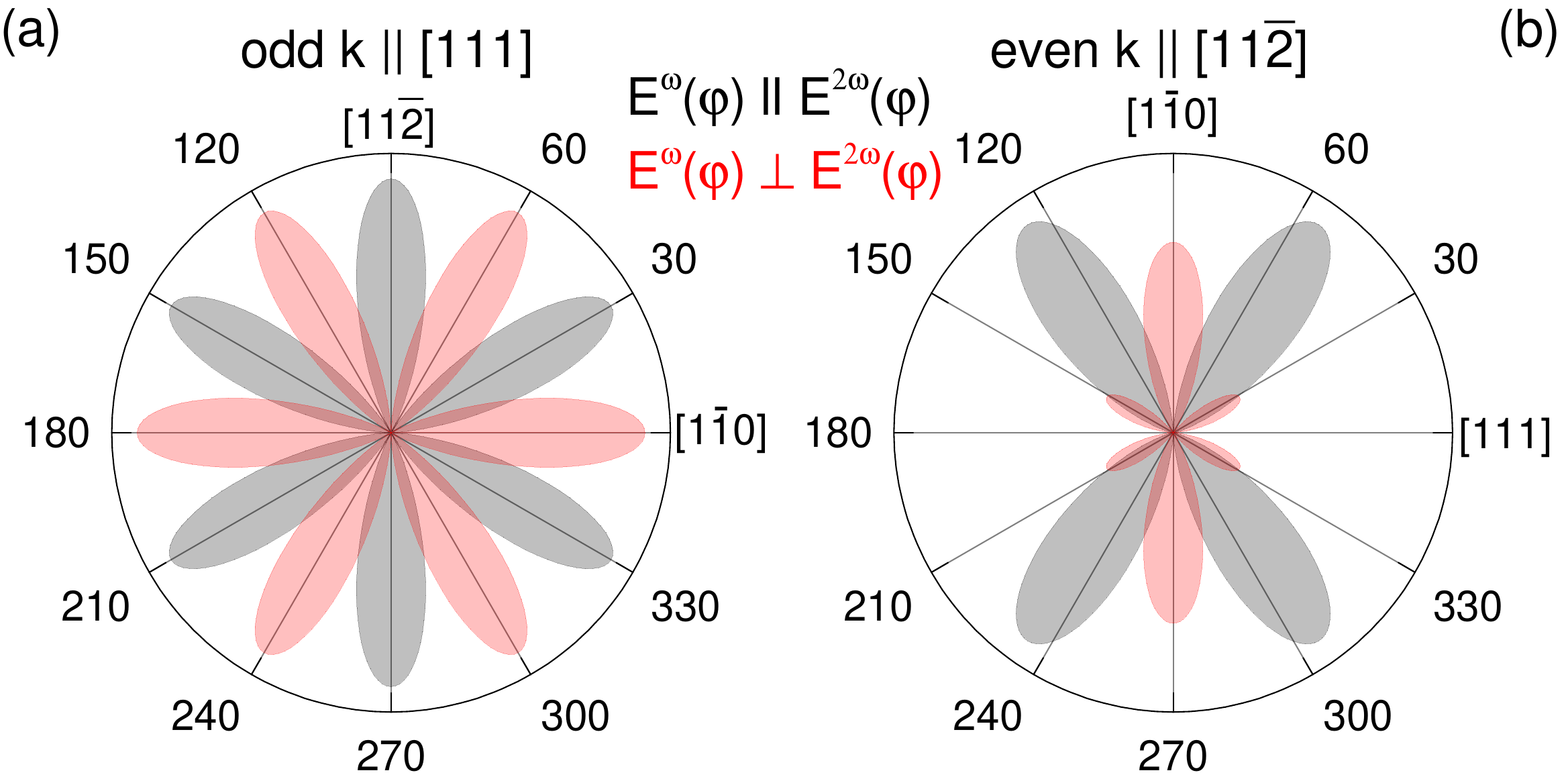}
\caption[Scan]{SHG rotational anisotropies in Cu$_2$O calculated for: (a) odd exciton states and $\textbf{k} \parallel [111]$ orientation, as well as (b) even exciton states and $\textbf{k}\parallel [11\bar{2}]$ orientation. Grey and red shaded areas represent the results for parallel and perpendicular polarization of the ingoing and outgoing photons, respectively.
}
\label{pic.Anisotropy_Simulation}
\end{center}
\end{figure}

The calculated rotational anisotropies are shown in Fig.~\ref{pic.Anisotropy_Simulation} for two configurations: (i) parallel linear polarizers: $\textbf{E}^\omega \parallel \textbf{E}^{2\omega}$ and (ii) perpendicular polarizers: $\textbf{E}^\omega \perp \textbf{E}^{2\omega}$.

\subsubsection{Anisotropies in experiment}

\begin{figure}[h]
\begin{center}
\includegraphics[width=0.5\textwidth]{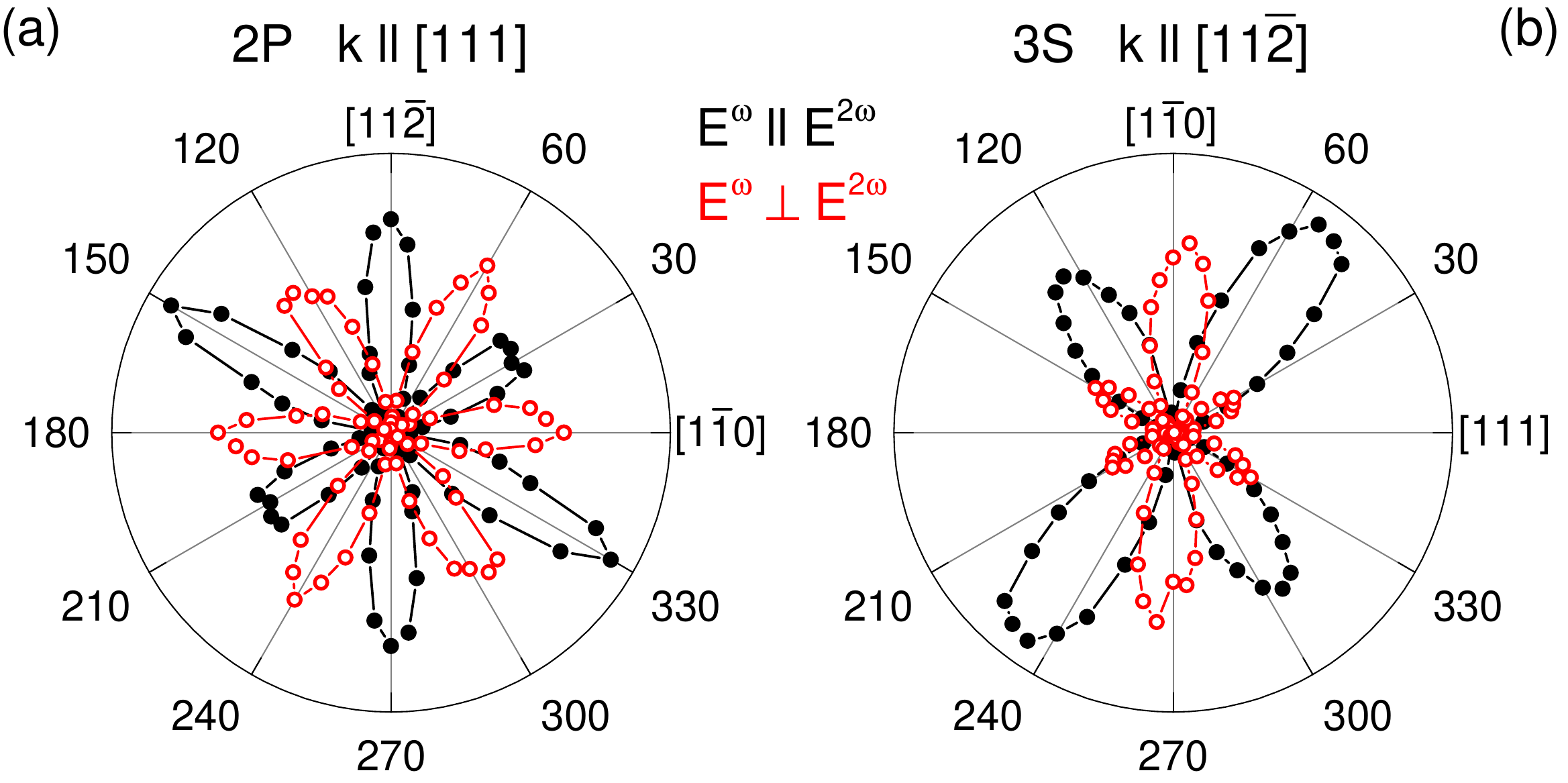}
\caption[Scan]{SHG rotational anisotropy of the $2P$ (energy 2.1475~eV, $\textbf{k} \parallel [111]$) and the $3S$ (2.1603~eV, $\textbf{k} \parallel [11\bar{2}]$) exciton states in Cu$_2$O. Filled (black) and open (red) circles represent data for the $\mathbf{E}^\omega \parallel \mathbf{E}^{2\omega}$ and the $\mathbf{E}^\omega \perp \mathbf{E}^{2\omega}$ configurations, respectively. 
}
\label{pic.Cu2O_Anis}
\end{center}
\end{figure}

 In Fig.~\ref{pic.Cu2O_Anis}(a) we show experimental data for the rotational anisotropies of the SHG signal considered before in the theory part. In particular, we show the rotational anisotropies for the $2P$ state in the $\mathbf{k} \parallel [111]$ orientation. Six-fold symmetries are seen both for the $\mathbf{E}^\omega \parallel \mathbf{E}^{2\omega}$ and the $\mathbf{E}^\omega \perp \mathbf{E}^{2\omega}$ configurations, which are rotated by 30$^\circ$ relative to each other. On the other hand, in Fig.~\ref{pic.Cu2O_Anis}(b) the $3S$ state measured for the $\mathbf{k} \parallel [11\bar{2}]$ orientation shows a four-fold symmetry in the $\mathbf{E}^\omega \parallel \mathbf{E}^{2\omega}$ and a six-fold symmetry in the $\mathbf{E}^\omega \perp \mathbf{E}^{2\omega}$ configuration.    
Comparing these data with the calculated diagrams from Fig.~\ref{pic.Anisotropy_Simulation} good agreement is clearly seen. The rather minor deviations are attributed to residual strain in the studied crystal.

Further, the excitons with $n>1$ are observed in SHG only when the laser light is sent in along crystallographic directions of reduced symmetry, e.g. for $\textbf{k}\parallel[111]$ and [11$\bar{2}$], while they are absent for $\textbf{k}\parallel[001]$ and [1$\bar{1}$0]. This is also in agreement with the theory, as one can see from Eqs.~\eqref{eq.SH_Oh_even} and ~\eqref{eq.SH_Oh_odd} the SHG signals vanish for $\textbf{k}\parallel[001]$ and [1$\bar{1}$0].

\section{Conclusions}

We have presented a novel, efficient way of second harmonic generation spectroscopy. By excitation using the broad band spectrum of fs laser pulses at half the energy of exciton resonances, narrow SHG lines are detected at energies corresponding to the exciton states, after spectral decomposition of the SHG signal. 

The efficiency of the method is based on the high peak power and the high repetition rate of the pulsed laser, leading to a good signal to noise ratio. This allows one to measure the SHG signal across a spectral range given by the width of the laser pulse without the need for tuning the photon energy. One has to be aware, that Sum-Frequency Generation (SFG) processes (excitation energies shifted by $\pm\Delta E$ from the SHG resonance) might contribute to the SHG signal at resonance. A profound analysis of these processes would probably have to take into account, that the fs-pulses result from phase-locked plain waves.

The spectral resolution is obviously limited only by the monochromator and the multi-channel detector, both of which may be improved compared to the current setting by using a monochromator with higher spectral resolution and a CCD with smaller pixel size.

We have applied the technique to the investigation of exciton states in Cu$_2$O bulk crystals. Besides observing the expected even parity $S-$ and $D-$excitons, we have also managed to detect odd parity excitons in SHG, not reported so far. These results challenged an analysis of the involved light-matter interaction mechanisms for which we used a group theoretical calculation. Our approach using only the relevant tables of Ref. \cite{Koster} is especially suited for semiconductor spectroscopy, where the exciton symmetries are easily derived from the band symmetries. For the more general approach starting with the relevant components of the nonlinear susceptibility tensor $\chi^{(2)}$ (third rank tensor) it is not so straight forward to implement quadrupole transitions, which lead to a fourth rank tensor. The predictions of this analysis for the polarization properties, for both linear and circular polarization, and for the rotational SHG anisotropies are confirmed by the experiments.

\begin{figure}[h]
\begin{center}
\includegraphics[width=0.45\textwidth]{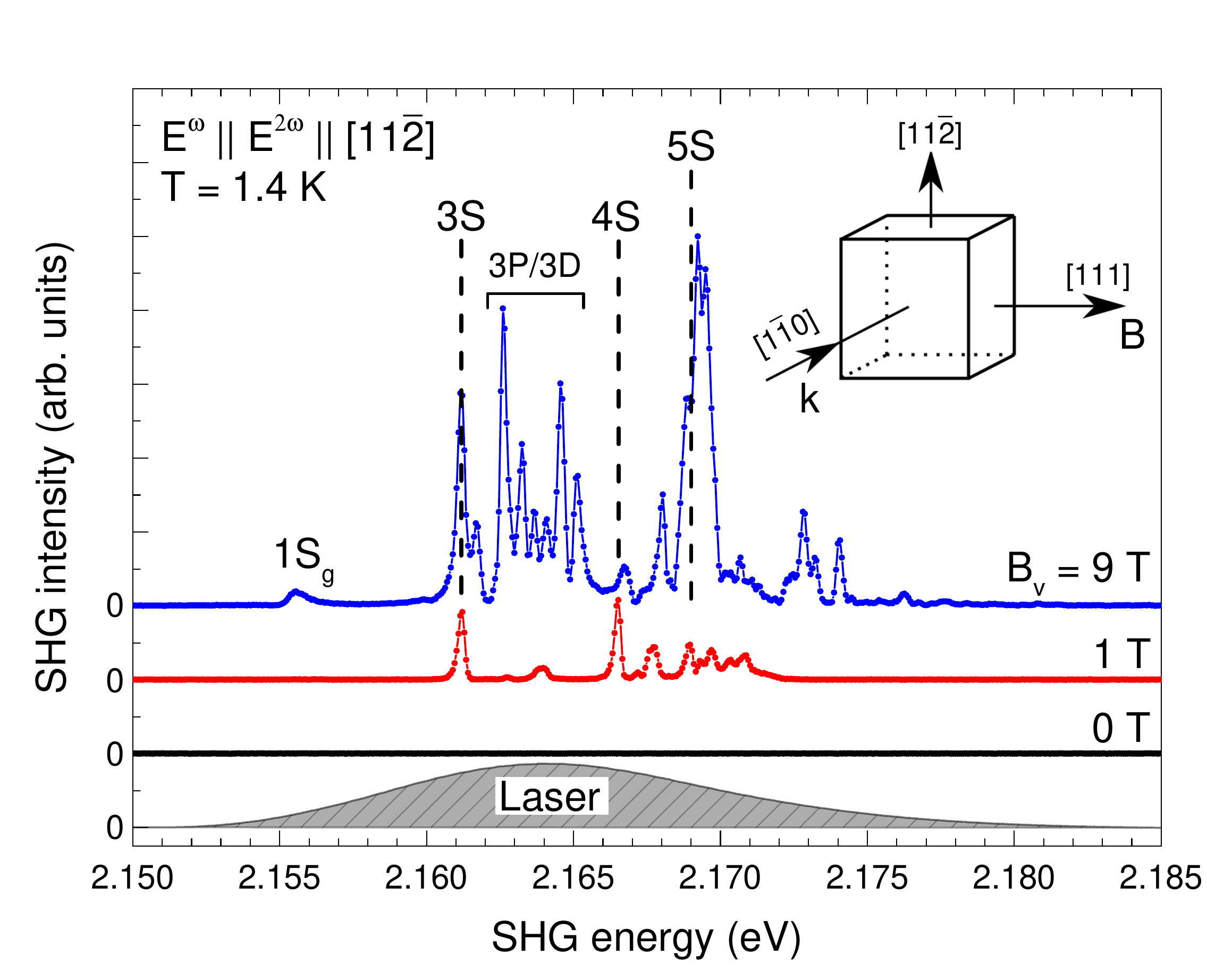}
\caption[Scan]{Magnetic-field-induced SHG spectra of Cu$_2$O in $\mathbf{k} \parallel [1\bar{1}0]$ orientation measured with linearly polarized light in the $\mathbf{E}^\omega \parallel \mathbf{E}^{2\omega} \parallel [11\bar{2}]$ configuration. Laser is set to 1.082~eV, its frequency doubled spectrum is shown by the grey shaded area. Magnetic field is applied in Voigt geometry.
}
\label{pic.1S_[111]_SHG_in_Bv+BBO}
\end{center}
\end{figure}

The potential of the method has been further assessed by first measurements in a magnetic field, using a configuration for which no exciton SHG is observed at zero field as seen in Fig.~\ref{pic.1S_[111]_SHG_in_Bv+BBO} for the $\mathbf{k} \parallel [1\bar{1}0]$ configuration. When applying magnetic fields up to 9~T in the Voigt geometry ($\mathbf{B} \parallel [111]$ and $\mathbf{B} \perp \mathbf{k}$), however, a rich set of SHG lines with strong intensities appears. Magnetic-field-induced SHG signals on exciton states have been reported for semiconductors such as GaAs, CdTe, (Cd,Mn)Te and ZnO \cite{Saenger06a,Saenger06b,Lafrentz13PRL,Lafrentz13,Brunne15}. Several possible mechanisms were identified for the field-induced SHG: the spin and orbital Zeeman effects and the magneto-Stark effect involving ED and EQ optical transitions. Further detailed studies are required to clarify which of these mechanisms are relevant for excitons in Cu$_2$O.

Obviously, the technique may be applied not only to bulk systems with pronounced exciton effects but also for  low-dimensional systems such as heterostructures and 2D materials, like dichalkogenides~\cite{Trolle2015,Wang2015,Glazov2017}. In fact, it is well suited for any SHG studies where high spectral resolution is needed, i.e. in solids, liquids, gases, and also biological objects, where SHG is used in microscopy for improving spatial resolution.

\begin{acknowledgments}
The authors are thankful to Ch.~Uihlein for stimulating discussions. We acknowledge the financial support by the Deutsche Forschungsgemeinschaft through the International Collaborative Research Centre TRR160 (Project C8) and the Collaborative Research Centre TRR142 (Project B01).
\end{acknowledgments}

\end{document}